\documentclass{article}
\usepackage{arxiv}
\usepackage{graphicx,color,amsmath,amsfonts,amscd,amssymb,pstricks,tikz}
\usepackage{dcolumn,tabularx}
\usepackage{bm}
\usepackage{siunitx}
\usepackage{multirow}
\usepackage{cite}

\usepackage{import}
\usepackage{xifthen}
\usepackage{transparent}

\usepackage{hyperref}

\title{Measurement of the nonlinear diffusion of the proton beam halo at the CERN LHC\thanks{Work funded by the HL-LHC project.}}

\author{C.E.~Montanari\\
University of Manchester and the Cockcroft Institute, Oxford Rd, Manchester M13 9PL, United Kingdom \\ 
and \\
Beams Department, CERN, Esplanade des Particules 1, 1211 Meyrin, Switzerland\\
\And 
R.B.~Appleby\\
University of Manchester and the Cockcroft Institute, Oxford Rd, Manchester M13 9PL, United Kingdom\\
\And 
A.~Bazzani\\
Dipartimento di Fisica e Astronomia, Universit\`a di Bologna, via Irnerio 46, 40126 Bologna, Italy \\ 
and \\
INFN sezione di Bologna, via Berti Pichat 6/2, 40127 Bologna, Italy\\ 
\And 
M.~Giovannozzi\thanks{Corresponding author: massimo.giovannozzi@cern.ch}\\
Beams Department, CERN, Esplanade des Particules 1, 1211 Meyrin, Switzerland\\
\And 
P.~Hermes\\ 
Beams Department, CERN, Esplanade des Particules 1, 1211 Meyrin, Switzerland\\
\And 
A.~Poyet\\
Beams Department, CERN, Esplanade des Particules 1, 1211 Meyrin, Switzerland\\
\And 
S.~Redaelli\\
Beams Department, CERN, Esplanade des Particules 1, 1211 Meyrin, Switzerland\\
\And 
G.~Sterbini\\
Beams Department, CERN, Esplanade des Particules 1, 1211 Meyrin, Switzerland}

\begin{document}
\maketitle

\begin{abstract}
In circular particle accelerators, storage rings, or colliders, mitigating beam losses is critical to ensuring optimal performance, particularly for rings that include superconducting magnets. A thorough understanding of beam-halo dynamics is essential for this purpose. This paper presents recent results for the measurement of the nonlinear diffusion process of the beam halo at the CERN Large Hadron Collider (LHC). The novel approach used in this paper is based on the analytical framework of the Nekhoroshev theorem, which provides a functional form for the diffusion coefficient. By monitoring the beam loss signal during controlled movements of the collimator jaws, we determine the beam losses at equilibrium for various amplitudes and analyze the beam-halo distribution. Post-processing of these measurements provides the nonlinear diffusion coefficient, which is found to be in excellent agreement with the theoretical assumptions. Measurements from an experiment investigating the effectiveness of beam-beam compensation using beam-beam compensation wires also provide a direct assessment of the compensation's effectiveness on beam-tail diffusion.
\end{abstract}


\section{Introduction}

In high-energy colliders and storage rings with superconducting magnets, beam dynamics is inherently complex and nonlinear due to magnetic field errors. These can cause beam losses or emittance growth, reducing accelerator performance and luminosity, and potentially leading to quenching of superconducting magnets, which affects operational efficiency.


Understanding nonlinear beam dynamics is crucial for predicting beam behavior, including beam losses and emittance growth. Although a relationship between dynamic aperture, the phase-space region where bounded motion occurs~\cite{DAdef,invlog,Bazzani:2019csk}, and beam lifetime has been established~\cite{da_and_losses}, it does not explain the evolution of the beam distribution, which is key for predicting emittance growth, a critical factor for evaluating collider performance.


To address this, a diffusive model framework, specifically the Fokker-Planck (FP) diffusion equation, offers a predictive approach for the beam distribution over long time scales (exceeding 10 hours), which is unachievable through direct particle tracking due to computational limitations. Current single-particle element-by-element tracking simulations for the LHC are restricted to only a few minutes of beam evolution (see, e.g.,~\cite{hermes:ipac2022-mopost045}).


The use of diffusive models to describe transverse beam dynamics is well established in accelerator physics (see, e.g.,~\cite{PhysRevLett.68.33,gerasimov1992applicability,MESS1994279,zimmermann1994transverse,PhysRevLett.77.1051,PhysRevSTAB.5.074001,stancari2011diffusion,PhysRevSTAB.15.101001}). However, recent advances~\cite{Bazzani:2019lse,bazzani2020diffusion,our_paper9} have introduced a new framework in which the functional form of the diffusion coefficient is derived from the optimal estimate of the perturbation series provided by the Nekhoroshev theorem~\cite{Nekhoroshev:1977aa,Bazzani:1990aa,Turchetti:1990aa}. As this approach offers a solid mathematical foundation for modeling beam behavior, it has the potential to significantly enhance our understanding of the processes behind beam losses and the formation of the beam halo. That is, a low-intensity, high-amplitude tail of the beam distribution, which deviates from the standard Gaussian form and is a significant source of beam losses.

The CERN Large Hadron Collider (LHC)~\cite{Bruning:782076} is a prime example where minimizing beam losses is crucial to maintaining optimal accelerator performance and hardware integrity. In the LHC, beam losses need to be safely absorbed by the collimation system to prevent the quenching of superconducting magnets and damage to any accelerator components~\cite{1590664, Assmann:972336, BRUCE201719, Bruce:1646958, Bruce:2686581}. Therefore, a thorough understanding of the beam-halo distribution and evolution is of particular interest. This challenge becomes even more relevant for future high-luminosity colliders, such as the High-Luminosity LHC (HL-LHC)~\cite{BejarAlonso:2749422, Arduini_2016}, where the increased beam intensity requires a precise assessment of expected beam losses to ensure their safe disposal~\cite{lindstrom:ipac2023-mopa126}. The situation will be even more challenging for future accelerators with higher stored energies, such as the CERN Future hadron Circular Collider~\cite{FCC-hhCDR,giovannozzi:ipac23-mopl033,ipac23fcchhcoll}.

This paper presents an innovative method for measuring the nonlinear diffusion coefficient governing the beam-halo dynamics that has been applied at the CERN LHC. By moving the collimator jaws following a recently proposed protocol~\cite{our_paper9,montanari:ipac2023-wepa022}, and observing the resulting beam loss signals, we determine the beam losses at equilibrium at various amplitudes and analyze the beam-halo distribution. These measurements allow us to accurately derive the nonlinear diffusion coefficient.

Furthermore, these measurements were performed with different configurations of Beam-Beam Compensator Wires (BBCW)~\cite{belanger:ipac2023-wepa060} on LHC Beam~2. BBCWs are devices used to mitigate the adverse effects of long-range beam-beam interactions (see, e.g., ~\cite{987147, Zimmermann:1955353, Rossi:2289671,  Fartoukh:2052448, Sterbini:2693922, axel.wires, belanger:ipac2023-wepa060, sterbini:ipac2023-wepl103} and references therein), which occur when particles in one beam exert electromagnetic forces on particles in the opposing beam. 
To compensate for these effects, BBCWs generate a magnetic field that counteracts the perturbations and are therefore expected to improve the beam lifetime without affecting the beam emittance. Our approach offers a useful tool for evaluating the diffusive behavior of beam dynamics in relation to the performance of BBCWs, providing direct insights into the effectiveness of these devices.

The structure of this paper is as follows. In Sec.~\ref{sec:diffusion}, we introduce the nonlinear diffusion model and the key physical observables used to measure the nonlinear diffusion. In Sec.~\ref{sec:method}, we describe the experimental method to measure the beam-halo distribution and the beam losses at equilibrium. In Sec.~\ref{sec:analysis}, we present the methodology used for the data analysis. In Sec.~\ref{sec:results}, we discuss the results and their implications. Finally, in Sec.~\ref{sec:conclusions}, we provide a summary and an outlook for future work. Some additional details on the critical aspects encountered during the measurement of LHC Beam~2 losses are presented in the Appendix~\ref{sec:b2}.

\section{nonlinear diffusion model}\label{sec:diffusion}

\begin{figure}[htp]
    \centering
    \includegraphics[width=0.8\columnwidth]{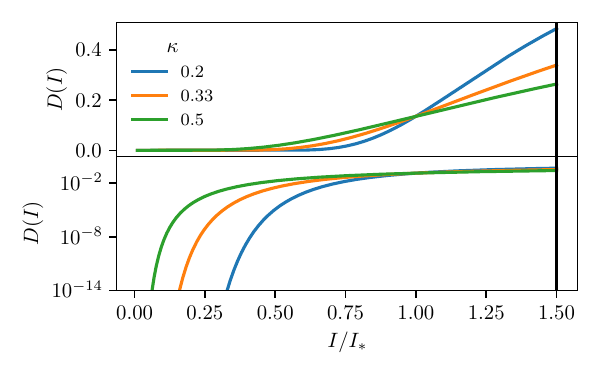}
    \caption{Overview of the functional form described in Eq.~\eqref{eq:d} for different values of $\kappa$, both in linear (top) and log scale (bottom).}
    \label{fig:functional_form}
\end{figure}

The modeling of the evolution of the transverse beam distribution by means of a diffusive framework is based on a rich foundation in accelerator physics. Within this framework, the FP equation
\begin{align}
    \frac{\partial \rho (I, t)}{\partial t} & =\frac{1}{2} \frac{\partial}{\partial I} {D}(I) \frac{\partial}{\partial I}\, \rho(I, t) \, , 
    \label{eq:fp}
\end{align}
along with the global diffusion coefficient $D(I)$ can provide a robust approach for determining the dynamics of the beam distribution, $\rho(I,t)$. Here, and in the rest of the article, $I$ denotes the action variable. Recent work \cite{Bazzani:2019lse,bazzani2020diffusion,our_paper9} offers significant progress in relation to previous approaches, linking the functional form of the diffusion coefficient with the optimal estimate derived from the Nekhoroshev theorem. This approach suggests a diffusion coefficient $D(I)$ that follows a Nekhoroshev-like functional form:
\begin{align}
    D(I) & \propto\exp \left[-2\left(\frac{I_\ast}{I}\right)^{\frac{1}{2\kappa}}\right]\, . 
    \label{eq:d}
\end{align}
This functional form depends only on $I$. The parameters $\kappa$ and $I_\ast$ have theoretical significance, relating to the analytic structure of the perturbative series and their asymptotic character, respectively \cite{bazzani2020diffusion}. A visualization of this functional form is presented in Fig.~\ref{fig:functional_form}. 

The boundary conditions of this FP system, which severely influence the solution of the system, consist of an absorbing wall at an amplitude $I_\mathrm{a}$, corresponding to the position of the collimator jaws in the accelerator machine at which the particles are lost, and a reflective boundary condition at $I=0$, although in this specific scenario where $\lim_{I\to 0^+} D(I)=0$ we are not affected by the typology of this boundary.

\subsection{Physical observables}


As we assume that the beam halo follows a diffusive behavior described by Eq.~\eqref{eq:fp}, with a diffusion coefficient $D(I)$~\eqref{eq:d} that becomes exponentially smaller for $I<I_\ast$, we can approximate the system by means of a virtual time-constant source term at position $I_0<I_\ast$ and a sink term placed at the position of the collimator jaw $I_{\mathrm{a}}$. This approximation requires the time scale for the evolution of the distribution at $I_0$ to be longer than the time scale for a measurement. By performing this approximation, we can treat the beam halo as a quasi-stationary system, where an analytical expression for the equilibrium distribution and the expected current loss can be achieved. 

With this hypothesis, we have that the beam halo quasi-stationary equilibrium distribution $\rho_{\mathrm{eq}}$ reads:
\begin{align}
    \rho_{\mathrm{eq}}\left(I , I_{\mathrm{a}}\right) & =\alpha\left(I_{\mathrm{a}}\right) \int_I^{I_{\mathrm{a}}} \frac{\mathrm{d} x}{D(x)}  \label{eq:equilibrium}\\
    \alpha\left(I_{\mathrm{a}}\right) & =\frac{\rho\left(I_0\right)}{\int_{I_0}^{I_{\mathrm{a}}} \frac{\mathrm{d} x}{D(x)}}\, , 
    \label{eq:equilibrium_alpha}
\end{align}
where $\rho\left(I_0\right)$ is the value of the beam distribution at the virtual source term, which is expected to be eroded over timescales longer than the duration of a typical measurement, allowing us to consider it as a constant for the duration of a measurement. The term $\alpha$ is then the constant factor that relates the equilibrium distribution to the source term, depending on the position of the absorbing boundary $I_{\mathrm{a}}$. Finally, from Eq.~\eqref{eq:equilibrium}, we can then derive the following expression for the beam losses at equilibrium
\begin{equation}
    J_{\mathrm{eq}}\left(I_{\mathrm{a}}\right)=\left.D\left(I_{\mathrm{a}}\right) \frac{\partial \rho_{\mathrm{eq}}(I)}{\partial I}\right|_{\left(I_{\mathrm{a}}\right)}=\alpha\left(I_{\mathrm{a}}\right) \, , 
    \label{eq:jeq}
\end{equation} 
which constitutes the key observable that we can measure with high reliability and precision. This observable corresponds to the equilibrium current as described in~\cite{our_paper9, montanari:ipac2023-wepa022}. Hereafter, we will refer to it using this terminology, which is typical of diffusion processes.

By performing an experimental measurement of the equilibrium current $J_{\mathrm{eq}}\left(I_{\mathrm{a}}\right)$ at different collimator-jaw positions and combining it with a measurement of the beam-halo distribution $\rho_{\mathrm{eq}}\left(I , I_{\mathrm{a}}\right)$, we can then fit the diffusion coefficient $D(I)$ to the nonlinear diffusion model described in Eq.~\eqref{eq:fp}. The fitting process will provide an estimate of the parameters $\kappa$ and $I_\ast$ that characterize the diffusion coefficient, and a choice of the parameter $I_0$ that characterizes the source term. 

\section{Experimental method}\label{sec:method}

\subsection{Specifics of the beam instrumentation}

The LHC provides a rich set of instruments for performing diagnostics and monitoring beam properties. Although a precise measurement of the beam-halo distribution, which exhibits both a vast dynamic range and a time dependence, is still an open challenge~\cite{Hermes:2023njr}, various beam observables are provided with appropriate precision and extended dynamic ranges, enabling a robust measurement of physical quantities related to beam-halo dynamics. All observables used in this work were recorded with the CERN Accelerator Logging System~\cite{Roderick:1215574} during the measurements and then processed and analyzed offline.

\subsubsection{Collimators}

The LHC collimation system is an essential component of machine operation and safety~\cite{1590664, Assmann:972336}. It serves fundamental purposes such as cleaning the beam halo, protecting the machine against unexpected and anomalous losses~\cite{BRUCE201719}, and reducing background noise in experimental regions~\cite{Bruce:1646958, Bruce:2686581}.

\begin{figure}[thp]
    \centering
    \includegraphics[width=0.6\columnwidth]{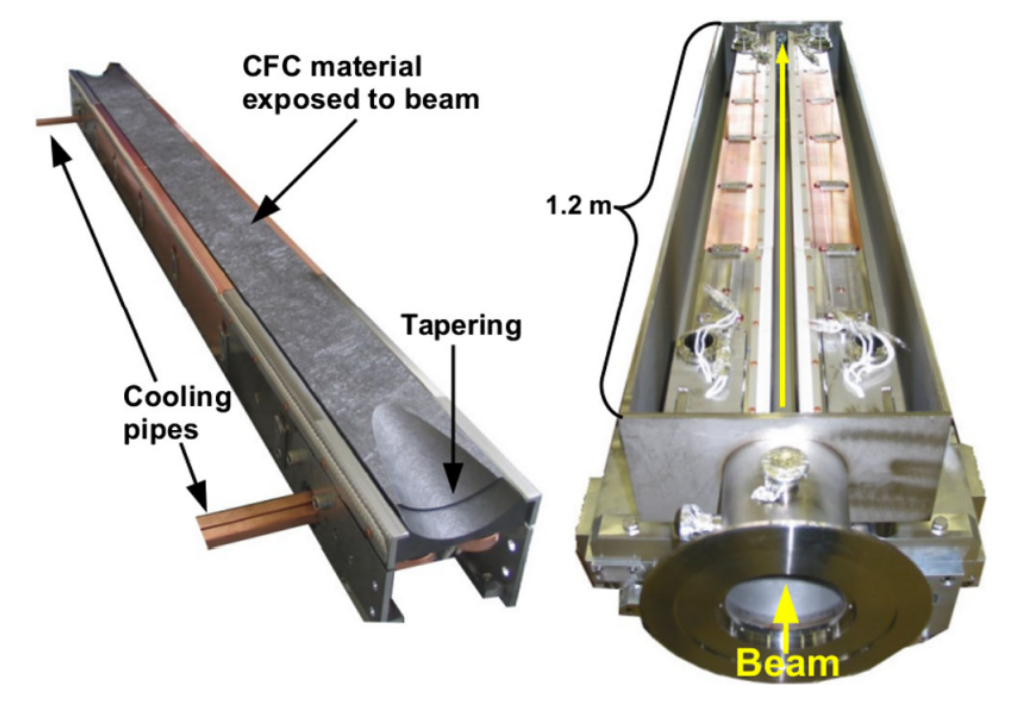}
    \caption{Left: jaw of a secondary collimator, made of carbon-fiber composite (CFC) and water cooled through copper pipes. Right: two collimator jaws are installed in a collimator tank (From Ref.~\cite{bruce2014simulations}).}
    \label{fig:collimator_pics}
\end{figure}

The collimation system consists of more than 120 individual collimators serving different purposes depending on their position and hierarchy. Most of these collimators are movable devices made up of two movable jaws, which can be carried at different distances from the circulating beam~\cite{Bertarelli:794628}. These jaws are straight, parallel to the beam and have a tapering at both ends along the beam axis to minimize the beam impedance. The distance between the start and end of the tapering, where the jaw material is straight, is referred to as the active length of the collimator. Some photos and schemes of these LHC collimators are reported in Fig.~\ref{fig:collimator_pics}. In the betatron cleaning insertion region of the LHC, IR7, we have a set of primary collimators, which are the collimators that define the physical size of the beam. These collimators are designed to intercept and absorb particles at an excessive amplitude and safely dispose of them. The resulting particle showers generated by such interactions are then measured by nearby sensors, which will be described in the following sections of the paper. Secondary collimators, tertiary collimators, and shower absorbers complete the multi-stage collimation system of the LHC. Primary halo particles interact with the primary collimators and are subsequently scattered to the secondary. The hadronic showers produced by the secondary collimators are ultimately absorbed by the shower absorbers, whereas the tertiary collimators protect the aperture bottlenecks of the triplet quadrupoles. 

The jaw positions and gaps of the primary collimators used during the experiment were monitored and recorded at a rate of \SI{1}{Hz} using Linear Variable Differential Transformers (LVDTs)~\cite{Masi:1110975}. The jaw motors of the collimator provide a minimal step of \SI{5}{\micro m}, which is the steps used for the collimator scans and scrapings.

\subsubsection{Beam losses}

The bulk of the data analysis is based on the observation of the beam losses along the LHC ring. These losses occur primarily in the IR7 region, where the losses induced by betatron cleaning are expected by design to be the most significant. These losses are mainly the result of electromagnetic and hadronic showers induced by the collimator jaws that intercept protons in the circulating beam. These losses are then detected and monitored by the Beam Loss Monitoring (BLM) system, which is composed of approximately 4,000 ionization chambers distributed along the LHC ring~\cite{blmSystem1}. The charges collected in these ionization chambers are then read out and processed over a dynamic range of about 8 orders of magnitude, ranging from a current of the order \SI{10}{pA} to \SI{1}{mA}. The final measurement is then provided in units of \si{Gy/s}, collected at a frequency of \SI{100}{Hz}~\footnote{For the sake of completeness, the LHC revolution period is approximately \SI{88.9}{\micro s}.}. The raw data are kept only for the most significant BLMs, such as those located in the IR7 region. All other BLM loss signals are provided at \SI{1}{Hz} in twelve different moving windows, also called running sums~\cite{rsdeftable, Dehning:895166}. In the running sums, the raw signal is combined in a fixed-duration moving window, acting as a low-pass filter, which effectively smooths out the signal and provides a less noisy observable. The resulting sum is then recorded at a rate of \SI{1}{Hz}, with a different protocol, depending on whether the time window is larger than the logging rate. For the entirety of this work, the so-called running sum RS09, corresponding to a time window of \SI{1310.72}{ms}, was used as the main observable for beam losses, as shorter time windows discard part of the signal due to the logging rate.

A calibration of the loss signal in \si{Gy/s} to the actual beam losses in \si{p/s} can be performed through dedicated BLM calibration measurements. During these measurements, controlled beam losses are induced in the jaws of different primary collimators, and the loss signal in BLMs is then correlated with the actual losses observed by other diagnostic tools, such as the Direct Current Beam Current Transformers (DCBCTs)~\cite{MoralesVigo:ibic21}, which are discussed next. This calibration provides a matrix of correlation factors that can be used to convert the loss signal from units of \si{Gy/s} to the actual losses in \si{p/s}. For the case of the measurements described here, the matrix based on the most recent calibration measurements was used. Note that BLM calibrations are performed using RS09 data.

Diamond beam loss monitors (dBLM) offer an extension of loss signal monitoring~\cite{Gorzawski:2320644, PhysRevAccelBeams.23.044802, calvogiraldo:ibic2022-tu2c2}, which enable a bunch-by-bunch measurement of the LHC bunch trains, each bunch being spaced by~\SI{25}{ns}, by providing a sampling time of \SI{1.6}{ns}. These detectors are based on the semiconductor properties of diamonds, where the ionizing particles generate electron-hole pairs that are then collected by the electrodes of the detector. Although a precise bunch-by-bunch calibration of the dBLM signal observed to the actual losses is not available at the time of writing, the data still offer a very useful overview of the relative losses between bunches and can be used to infer the presence of single-bunch instabilities or other effects that cannot be discerned from the standard BLM signal. 

\subsubsection{Beam intensity}

DCBCTs~\cite{Denard:1213275} provide a measurement of the total intensity of the beam that circulates around the LHC circumference. These transformers, consisting of toroidal designs with multiple windings, detect the passage of charged particles by inducing currents proportional to beam intensity. The measurement is provided in the number of protons and is recorded at a rate of \SI{1}{Hz}. The BCT signal is used to monitor the evolution of the beam intensity during the collimator scans and scrapings, and assesses the global intensity evolution of the beams during beam operations.

A bunch-by-bunch measurement of the beam intensity is provided by the Fast Beam Current Transformers (FBCT)~\cite{Belohrad:1267400}, and is used to monitor the evolution of the beam intensity during collimator scans and scrapings. The logging of the FBCT signal is performed at a rate of \SI{1}{Hz}, and provides a calibrated measurement in the number of protons, similarly to DCBCT. Compared to the loss signal observed by the BLMs, we have that both the DCBCT and FBCT signals are not sensitive enough to detect the small and fast variations in the beam loss rates induced by collimator movements, except for significant scraping events characterizing dedicated calibration measurements. However, the FBCT signal can be used to observe the overall evolution of the beam intensity during the measurements and to detect anomalous behaviors of individual bunches. 

\subsubsection{Beam emittance}

A measurement of the beam emittance is provided by means of the Beam Synchrotron Radiation Telescope (BSRT). The BSRT is based on the observation of the synchrotron radiation emitted by the beam, which is then collected by a set of optical elements and focused on a detector. The detector is then read out and processed to provide an estimate of the beam emittance. In this analysis, the measured beam emittance is used to convert the units from millimeters to beam sigma units.

\subsection{Machine configuration}

The measurement was carried out at the End of Fill (EoF) of a machine development experiment designed to study the effectiveness of beam-beam compensation in the LHC with the BBCW. This study, during fill 8348, took advantage of bunches more intense than those in standard Run~3 operation, a lower crossing angle, and a modified filling scheme to reduce the effect of the electron cloud~\cite{Iadarola:2804500}. More specifically, a filling scheme with 48 bunches per train (that is, 48 bunches separated by \SI{25}{ns}) was used, with a total of 3 trains, distanced as much as possible. 14 non-colliding bunches were also present in the machine to provide a reference sample not affected by the beam-beam interactions. Note that BBCW were available for Beam~2 only, as the wires in Beam~1 were not operational. The layout of the BBCW consists of 2 copper wires per beam and per interaction point (IP) around IP1 and IP5 (location of the ATLAS and CMS experiments) embedded in the jaws of special collimators. The wires are placed in the vertical plane for IP1 and in the horizontal plane for IP5. 

Table~\ref{tab:beamparameters} summarizes the key beam parameters that highlight the differences between the experiment and a standard operational case (e.g., fill 8120), the latter providing a baseline for Run~3 observations. During the experiment, the half-crossing angle was reduced from \si{160} to \SI{130}{\micro rad} to investigate the effectiveness of beam-beam compensation at different intensities of long-range beam-beam effects. A complete overview of the machine configuration and the experimental setup can be found in Ref.~\cite{belanger:ipac2023-wepa060}. When the EoF measurements were made, the crossing angle was set to \SI{130}{\micro rad}.

\begin{table}[thp]
\centering
    \caption{\label{tab:beamparameters}%
    Comparison of key beam parameters between the EoF 8348 and a standard operational case used in Run~3~\cite{belanger:ipac2023-wepa060}. The wire position is given with respect to the nominal reference orbit.}
        \begin{tabular}{lcccl}
        \hline
        \textbf{Parameter}& &
        \textbf{Exp.}& \textbf{Oper.}& \textbf{Unit} \\
        & & & \textbf{Fill} & \\ \hline
        Wire current & $I_\mathrm{W}$ & 350 & 350 & (A) \\
        Wire pos.\ IP1$|$
        IP5 & $d_{W}$ & $9.2|12.4$ & $9.2|12.4$ & (mm) \\
        Beam energy & $E$ & 6.8 & 6.8 & (TeV) \\
        Bunch Intensity & $N$ & $1.4$ & $1.0$ & ($10^{11}$ ppb) \\
        Beta at IPs & $\beta^*$ & 30 & 30 & (cm) \\
        Half-crossing angle & $\theta_c/2$ & 130 & 160 & (\si{\micro rad}) \\
        Num.\ of bunches & $n_b$ & 158 & 2413 & \\
        Bunches per train &  & 48 & 48 & \\ \hline
        \end{tabular}
\end{table}

\subsection{Experimental procedure}

An experimental measurement of the beam-halo distribution $\rho_{\mathrm{eq}}\left(I , I_{\mathrm{a}}\right)$ is obtained by means of a scraping procedure, with the primary collimator jaw starting from $I_{\mathrm{a}}$, during which the beam halo is removed by the collimator jaw by consecutive steps inward, and the observed beam losses are integrated to reconstruct the beam-halo distribution. This measurement constitutes a standard technique in the context of LHC operations, as can be seen, e.g., in Ref.~\cite{Burkart:2011lrs, Valentino:1480603, FusterMartinez:2019bqo, PhysRevAccelBeams.23.044802, Hermes:2023njr}. The first collimator scraping also serves the purpose of removing the beam-halo population generated either in the injectors and then injected in the LHC or during the previous operational LHC stages, which might have overpopulated the halo due to the multiple changes in the machine configuration. These conditions are not representative of the beam-halo distribution generated by the diffusion process that takes place during the collimator scan stages. 

\begin{figure}[thp]
    \centering
    \includegraphics[width=0.8\columnwidth]{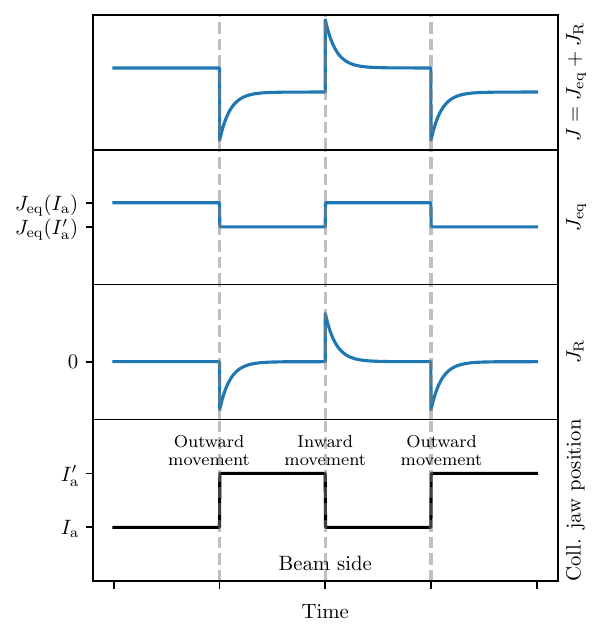}
    \caption{Schematic representation of the current loss signal $J$ (first plot) observed during a three-step movement of the collimator-scan procedure (fourth plot). $J$ is the sum of two components $J_{\text{eq}}$ (second plot) and $J_{\mathrm{R}}$ (third plot).}
    \label{fig:scan_schema}
\end{figure}

Regarding the measurement of the equilibrium current $J_{\mathrm{eq}}\left(I_{\mathrm{a}}\right)$, we follow the collimator-scan procedure proposed in Ref.~\cite{our_paper9}, which consists of alternating movements of the collimator-jaw positions following a three-step movement. That is, a outward-inward-outward movement with long enough pauses between steps to allow the system to relax to the new equilibrium. By performing this procedure, we can isolate two key components of the current loss signal, $J(t)$, namely, the equilibrium current $J_{\text{eq}}(t)$, and the recovery current $J_{\mathrm{R}}(t)$. The equilibrium current $J_{\text{eq}}(t)$ is related to the slow erosion of the beam core, while the recovery current $J_{\mathrm{R}}(t)$ constitutes an out-of-equilibrium signal given by the system when a rapid adjustment occurs from a quasi-equilibrium state to a new one, due to a change in the collimator-jaw position. A sketch of the collimator scan and the two components of the current loss signal is shown in Fig.~\ref{fig:scan_schema}.

During our measurements, a combination of collimator scrapings and collimator scans were performed twice on the horizontal plane of Beam~1 and Beam~2 with one of the jaws of the IR7 primary collimators. The procedure was performed first with the Beam~2 BBCW on, and then with the Beam~2 BBCW off. Before the beginning of the first measurement, a beam-based alignment~\cite{valentino:ipac12-tuppr098} was performed to assess the position of the center of the beam at IR7. The measurements were performed in parallel on the two beams.

For each measurement, an initial beam scraping was performed to remove the beam halo to the smallest aperture compatible with the beam loss safety thresholds. 
This scraping and subsequent scrapings were performed by moving the collimator jaw inward in steps of \SI{5}{\micro m}, corresponding to a movement of $\simeq0.025 \sigma$ in Beam~1 and of $\simeq0.023 \sigma$ in Beam~2, with a pause between steps to avoid excessive pile-up of beam losses. These pauses ended up being on the order of \SI{10}{s} for the first and most intensive scraping, and of the order of \SI{5}{s} for the subsequent ones.

After this initial scraping, the collimator scan is performed, with the jaw moving in steps of \SI{5}{\micro m}, and pauses between steps of the order of \SI{60}{s}. Scans were performed in the same region of amplitude as explored by scraping.

At the end of the scan, since the beam halo is expected to be in the quasi-stationary equilibrium state described in Eq.~\eqref{eq:equilibrium}, a final scraping is performed to probe the beam-halo distribution. This final scraping also provides the information to infer the value of $\rho(I_0)$.

\begin{figure*}[ht]
    \includegraphics[width=\textwidth]{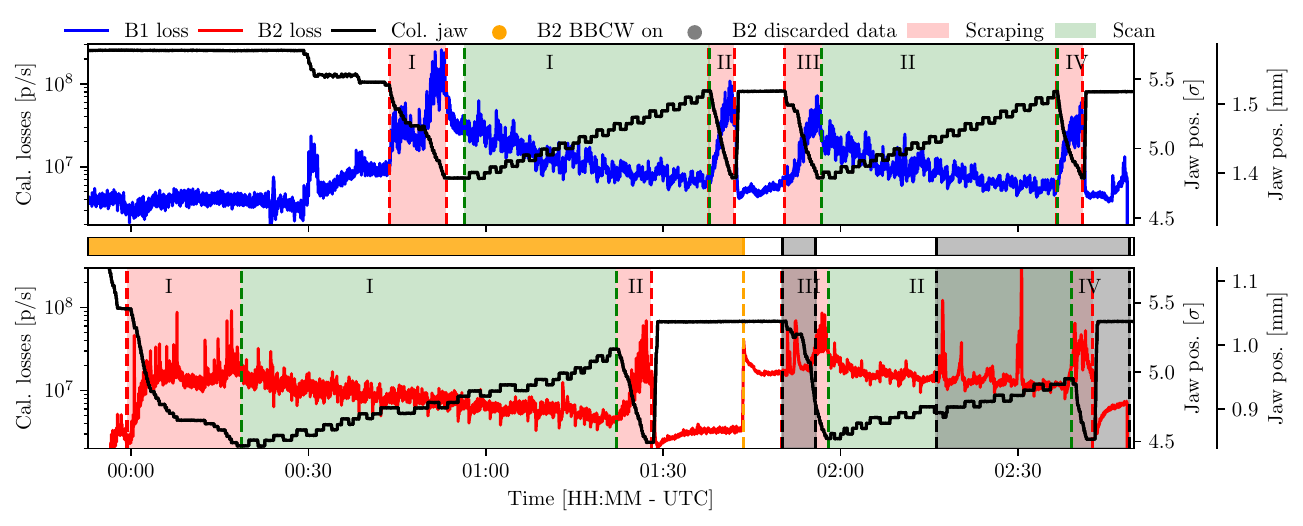}
    \caption{Overview of the collimator-jaw positions set and beam losses observed for Beam~1 (top) and Beam~2 (bottom) for the overall duration of the measurements. Highlights showing the corresponding time windows where scrapings (red) and scans (green) were performed are also shown. The central colored rectangle highlights the time windows where the BBCW was on (yellow) and off (white) on Beam~2, with the vertical orange dashed line in the bottom plot highlighting the moment when the BBCW was turned off. The large spikes observed in the Beam~2 losses with the BBCW off are related to single-bunch instabilities, that were consequently discarded (gray highlights).}
    \label{fig:collimator_scan}
\end{figure*}

In Fig.~\ref{fig:collimator_scan} we show an overview of the collimator-jaw positions for Beam~1 and Beam~2 for the overall duration of the measurements. The figure shows the jaw positions for the first and second measurements, with BBCW on and off, respectively. The calibrated losses observed from the BLM system are also shown. It can be immediately observed that the losses in Beam~2 with the BBCW off manifest some significant spikes that are absent in the other measurements. These spikes are related to single-bunch instabilities that affected the measurements of Beam~2.

\section{Data analysis}\label{sec:analysis}

\subsection{Measurement of the equilibrium current}

The equilibrium current $J_{\mathrm{eq}}$ was measured at different collimator-jaw positions $I_\mathrm{a}$ during the various steps of the collimator scan. Although it was possible to observe a clear stabilization of the beam loss signal at each step, after a waiting time of the order of \SI{15}{s}, the signal was observed to be noisy and affected by occasional spikes, most likely related to spurious effects, not related to the diffusion dynamics under investigation, such as orbit jitters.

\begin{figure}[thp]
\centering
    \includegraphics[width=0.8\columnwidth]{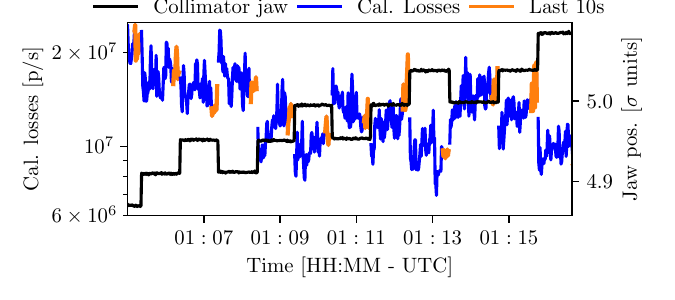}
    \caption{Detail of the equilibrium current measurement for Beam~1. The equilibrium current is measured as the average of the signal over a time window of \SI{10}{s}, taken right before the next collimator-jaw movement. The uncertainty of the measurement is considered as the standard deviation of the signal computed over the same time window.}
    \label{fig:scan}
\end{figure}

To mitigate these effects and associate an uncertainty with the measurement, the equilibrium current was measured for each collimator scan step as the average of the BLM calibrated signal over a time window of \SI{10}{s}, taken before the next collimator-scan step. For the uncertainty of this measurement, we considered the standard deviation of the signal over the same time window. A graphical visualization of the measurement method is shown in Fig.~\ref{fig:scan}, for a selection of Beam~1 data. 

The equilibrium current for Beam~1 was measured in two separate collimator scans. These two collimator scans were executed in the same plane under the same machine configuration, with a time difference of around one hour of machine time with colliding beams. The trend observed for the two measurements shows that the equilibrium current decreases as the collimator jaw is set at a higher amplitude. Compared, the two observations show a constant offset, with the second equilibrium current being around $20\%$ smaller than the first. This offset can be attributed to the gradual erosion process described above and can be modeled with the term $\rho(I_0)$, expected to slowly decrease over long timescales. As will be shown later, the evolution of this term can be inferred from the fitted parameters of the diffusion coefficient.

As for Beam~2, the equilibrium current was measured initially with BBCW on and then with BBCW off. A significant difference is observed between the two measurements, with the equilibrium current being significantly higher when the BBCW is off. 

\subsection{Reconstruction of beam-halo population}

To obtain an estimate of the beam-halo population $\rho_\mathrm{eq}$, we considered the calibrated beam loss signal observed during the collimator scraping measurements. The measurement is based on the assumption that the scraping is performed fast enough to observe only beam losses that are due to the beam halo removed by the collimator jaw. The losses observed can then be integrated over the duration of the various inward movements of the collimator jaw to reconstruct the number of particles that populate the amplitude interval scraped during that specific movement. By analyzing the integrated losses observed at each step, we can infer the beam-halo distribution.

\begin{figure}[thp]
\centering
    \includegraphics[width=0.7\columnwidth]{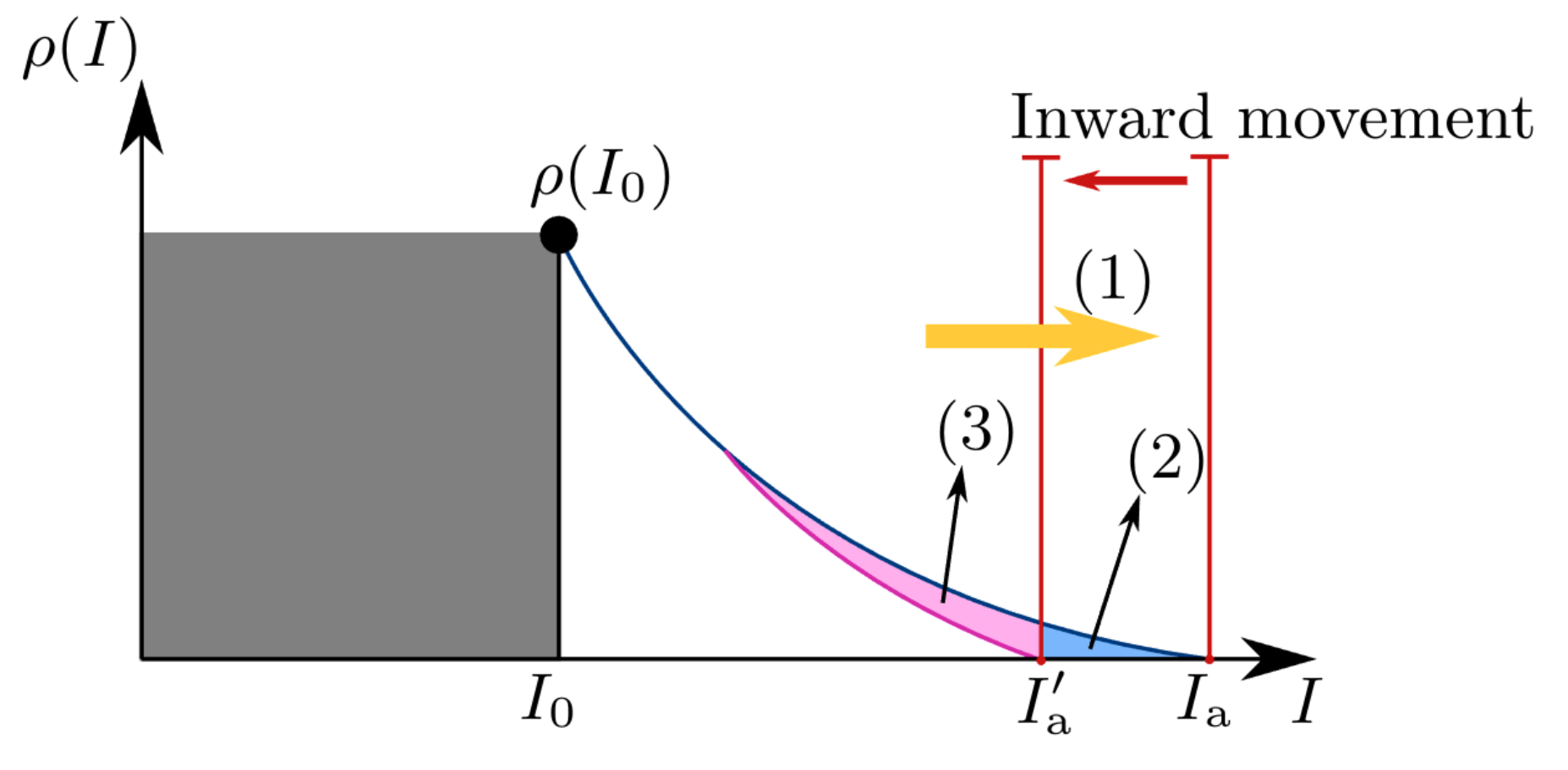}
    \caption{Schematic representation of the three processes contributing to the observed loss signal during collimator scraping inwards movement: (1) the underlying equilibrium current $J_{\mathrm{eq}}$; (2) the beam-halo population that is removed by the collimator jaw; (3) the out-of-equilibrium relaxation process of the remaining beam-halo population.}
    \label{fig:scraping_schema}
\end{figure}

In the context of a collimator-jaw scraping, the observed beam loss signal can be considered as the sum of three different processes: (1) the underlying equilibrium current $J_{\mathrm{eq}}$ that is expected to be observed at the starting position of the collimator jaw; (2) the beam-halo population that is expected to be removed by the collimator jaw; (3) the out-of-equilibrium relaxation process of the remaining beam-halo population that will occur due to the new equilibrium condition imposed by the new jaw position. A schematic representation of the three processes is shown in Fig.~\ref{fig:scraping_schema}.

Although the first process is directly observed and measured by the collimator-scan process and can therefore be directly subtracted from the observed beam loss signal, there is no straightforward way to separate the second process from the third, as the out-of-equilibrium relaxation does not offer a clear signature in the loss signal. However, since full equilibrium relaxation was observed to occur on a time scale of the order of \SI{10}{s} along the steps performed at different amplitudes in the collimator scan, we can assume that the scraping process is fast enough to remove the beam halo before any significant relaxation occurs, as collimator scrapings after scans were performed in steps of approximately \SI{3}{s}.

\begin{figure*}[thp]
    \includegraphics[width=\textwidth]{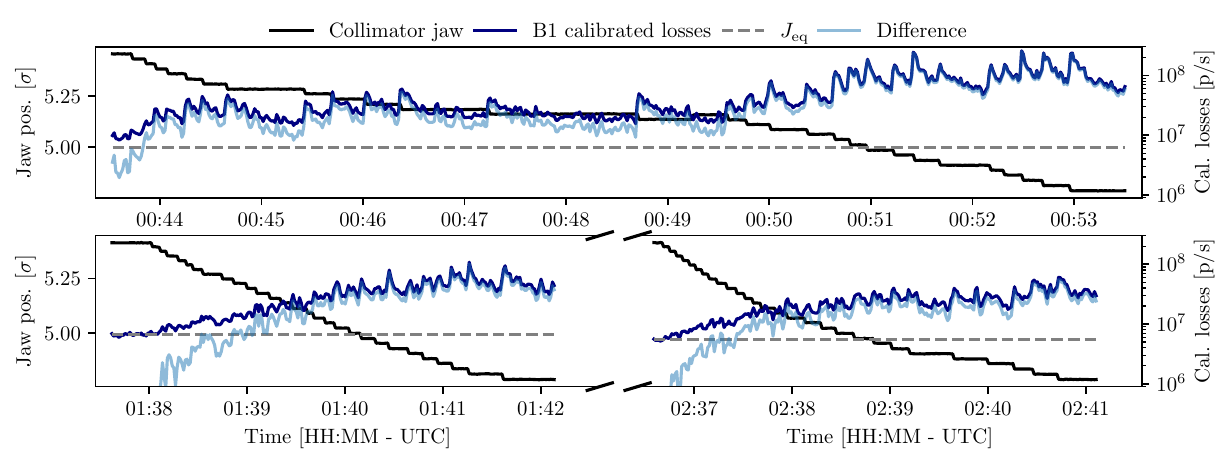}
    \caption{Overview of the observed calibrated losses on Beam~1 during the first (upper plot), the second (lower-left plot), and fourth (lower-right plot) scraping measurement. The value of $J_{\mathrm{eq}}$ observed during the collimator scan at the amplitude corresponding to the initial position of the collimator jaw, observed during the corresponding collimator scan, is also shown. The first scraping displays a more populated beam halo than the other two measurements.}
    \label{fig:scraping-b1}
\end{figure*}

Figure~\ref{fig:scraping-b1} shows the losses observed during the first, second, and fourth scraping of Beam~1, along with the value of $J_{\mathrm{eq}}$ observed during the collimator scan at the amplitude corresponding to the initial position of the collimator jaw. The observed beam losses are then corrected for by subtracting the value of $J_{\mathrm{eq}}$. It can be seen that the first scraping shows signal peaks 3-5 times higher than those observed in the second and fourth scrapings, which are conversely closely comparable. This difference reflects the different conditions of the beam-halo population observed during the first scraping and highlights the importance of relying on the measurements performed right after the collimator scans to provide a reliable estimate of the underlying diffusion process.

\begin{figure}[thp]
\centering
    \includegraphics[width=0.8\columnwidth]{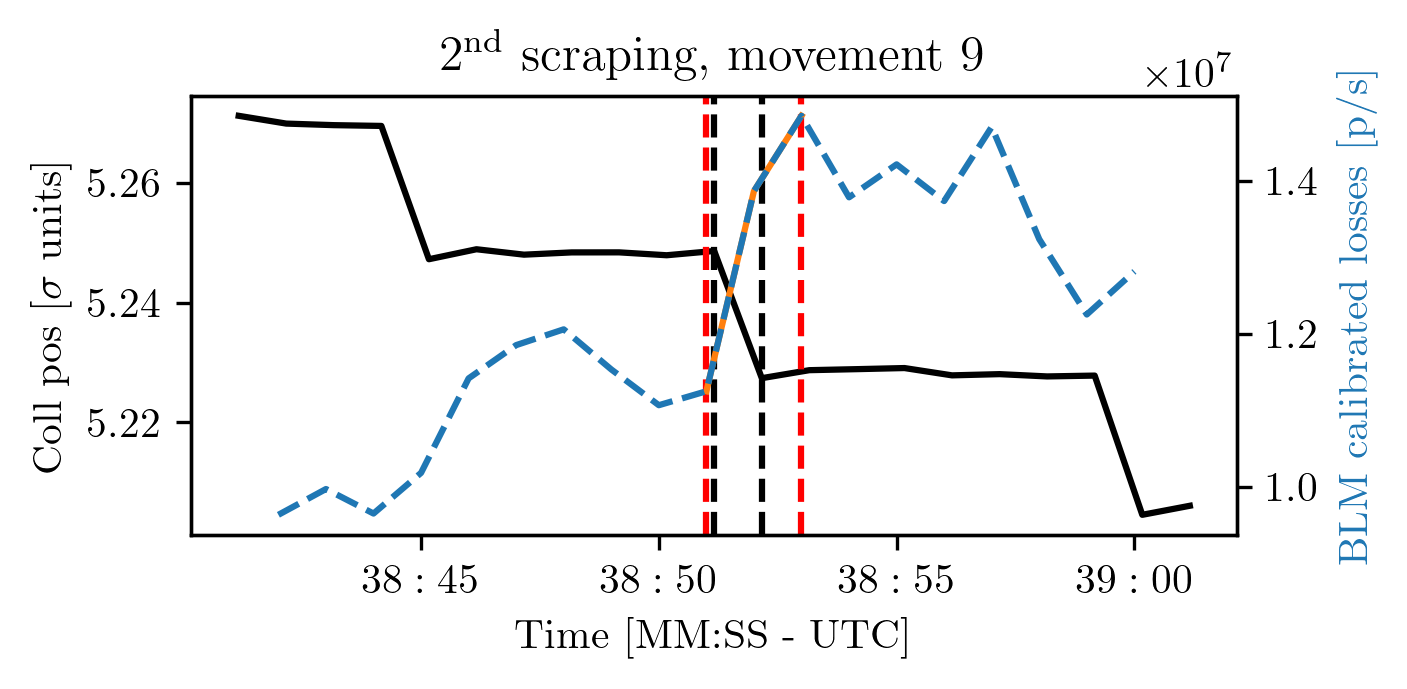}
    \includegraphics[width=0.8\columnwidth]{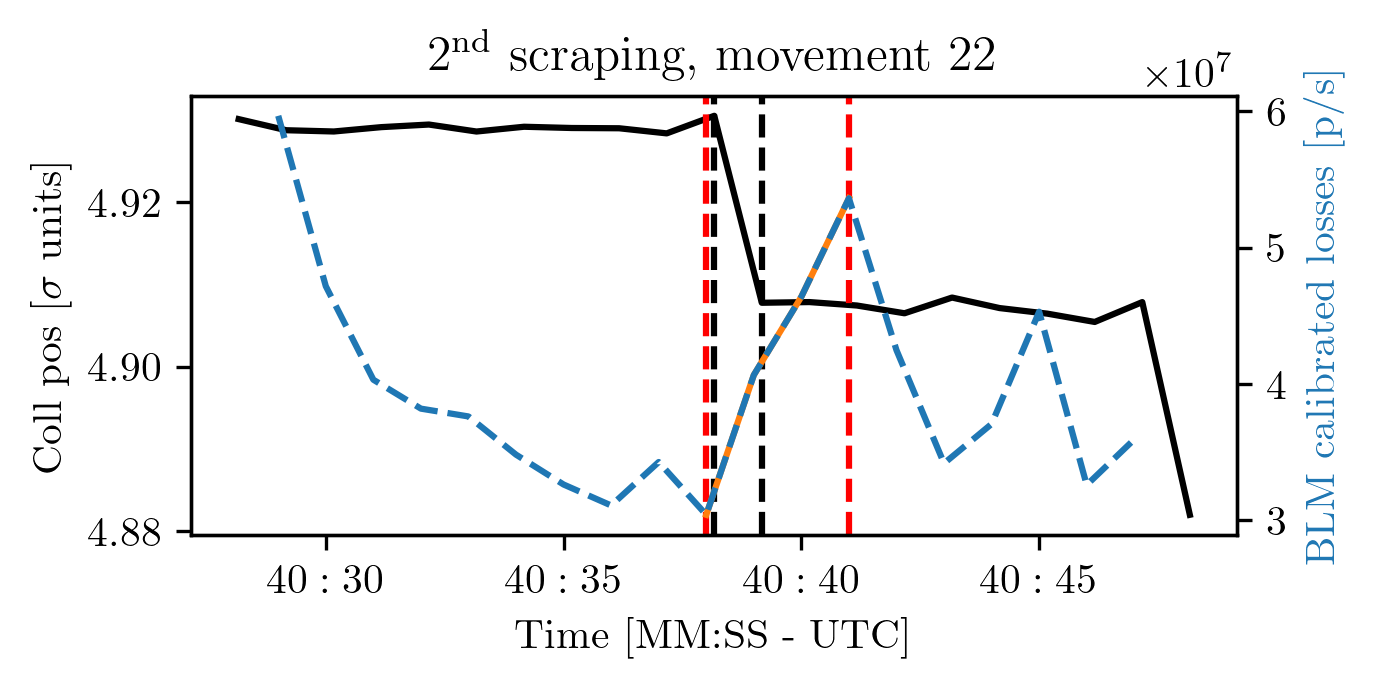}
    \caption{Detail of the observed beam losses during two inward movements of the collimator jaw performed in the second scraping measurement of Beam~1. The black dashed lines highlight the time interval in which the collimator jaw was moving, while the red dashed lines highlight the time interval selected for the integrated loss measurement, according to the selection method used.}
    \label{fig:scraping-b1-detail}
\end{figure}

A zoom in on some individual inward movements of the collimator jaw for the second scraping and the corresponding losses are shown in Fig.~\ref{fig:scraping-b1-detail}. It can be observed that the peak in the loss signal is not always perfectly synchronized with the inward movement of the collimator jaw, suggesting a certain loss of synchronization of the two measurements. To address this issue, instead of considering only the losses observed during the collimator movement, we considered the full loss signal increase observed near the collimator movement, within a time window of an extra second before and after the collimator movement. Within this time window, an increase in the beam loss signal is detected. The resulting selected intervals following this approach are also shown in Fig.~\ref{fig:scraping-b1-detail}. Note that the selected interval in the loss signal does not overlap with the beginning and end of the measured collimator-jaw movement. As an uncertainty of the integrated measurement, we consider the relative difference between the integral over the chosen interval and the integral over the interval increased by half a second on each side, applying a linear interpolation of the data points. This choice was made to account for the possible uncertainty in the synchronization of the two measurements and due to the \SI{1}{Hz} logging of the data. 

\begin{figure}[thp]
\centering
    \includegraphics[width=0.8\columnwidth]{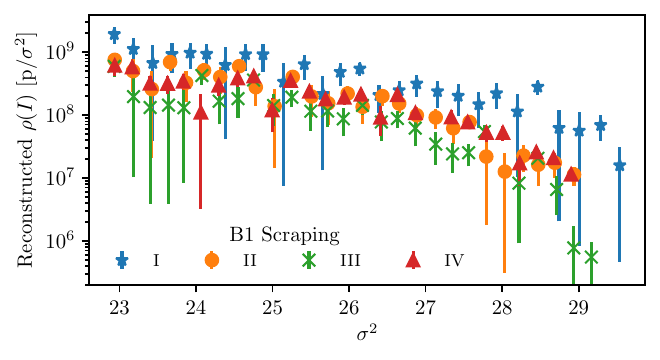}
    \caption{Comparison of the reconstructed proton density distribution from the scraping measurements performed on Beam~1. The first scraping measurement presents a higher beam-halo population than the other three measurements, which are close to comparable. The third scraping, executed a few minutes after the second one, shows a slightly lower beam-halo population. The second and fourth scrapings are close to comparable.}
    \label{fig:scraping-b1-result}
\end{figure}

The four scraping measurements are presented in Fig.~\ref{fig:scraping-b1-result}, where it can be observed that the first scraping exhibits a higher beam-halo population than the other three measurements, which appear to be closely comparable, considering the uncertainties of the measurements. The third scraping, while close to the second and fourth ones in magnitude and comparable if considering the experimental error, still exhibits a slightly lower beam-halo population, in line with the shorter time elapsed from the second scraping.

\begin{figure}[thp]
\centering
    \includegraphics[width=0.8\columnwidth]{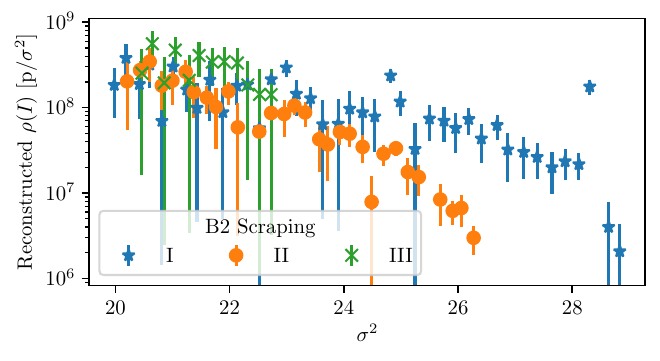}
    \caption{Comparison of the reconstructed proton beam density distribution from the scraping measurements performed on Beam~2. The fourth scraping measurement is not shown, as it was affected by single-bunch instabilities discussed in Appendix~\ref{sec:b2}. The third one only considers the fraction of data available.}
    \label{fig:scraping-b2}
\end{figure}

Regarding the scraping measurements performed on Beam~2, we can consider only the first two scrapings and parts of the third, while all data collected from the fourth scraping had to be discarded. The reconstruction obtained from the valid scraping measurements is presented in Fig.~\ref{fig:scraping-b2}. It can be observed that a similar trend in $\rho_\mathrm{eq}(I)$ is observed in the first two scrapings, the second showing a slightly lower beam-halo population. The reconstruction of $\rho_\mathrm{eq}(I)$  obtained from the second scraping will be used along the first scan data to reconstruct the diffusion coefficient for Beam~2 with BBCW on, following an approach similar to that used for Beam~1. The third scraping will be considered, instead of the fourth, for the diffusion fit of Beam~2 with BBCW off, together with the second collimator scan data. Although this last choice is not ideal and is not fully in line with the methodology used for Beam~1 and Beam~2 with BBCW on, this third scraping offers the only option to retrieve at least an estimate of the intensity of the beam-halo population at low enough amplitudes to be able to determine $\rho(I_0)$ for the case with BBCW off.

\subsection{Determination of the diffusion coefficient}

The equilibrium current measurements $J_\mathrm{eq}$ and the beam-halo population reconstructions $\rho_\mathrm{eq}$ have been used to fit the diffusion coefficient $D(I)$ in the functional form of Eq.~\eqref{eq:d}, with $I_\ast$ and $\kappa$ as free parameters. To fit the beam-halo population data and the equilibrium current data, Eqs.~\eqref{eq:equilibrium} and~\eqref{eq:jeq} are used, respectively, requiring the addition of $I_0$ as a third free parameter, representing the point at which we treat the beam-halo population as a virtual source term. The intensity of the source term $\rho(I_0)$ is then inferred from the beam-halo distribution obtained from the scraping measurements, as described in the previous section.

The fit was performed by minimizing the weighted sum of the squared residuals between the observed equilibrium current and beam-halo population with the model predictions of these two quantities. The residuals were weighted by the uncertainties of the measurements. As the two measurements consider two different physical quantities that range over different orders of magnitude and different signal-to-noise ratios, a multiplicative factor $\eta$ was introduced for the residuals from the beam-halo population and varied with the goal of analyzing the different fit results obtained. The $\chi_{\mathrm{total}}^2$ was defined as
\begin{equation}
    \chi_{\mathrm{total}}^2 = \chi_{J_{\mathrm{eq}}}^2 + \eta \chi_{\rho}^2\, , \label{eq:residuals}
\end{equation}
where $\eta$ is varied logarithmically between $10^{-2}$ (most significance to the scan data and $J_\mathrm{eq}$) and $10^{2}$ (most significance to the scraping data and $\rho_\mathrm{eq}$) to inspect the different fit results and assess the overall robustness of the measurements. When $\eta=1$, the residuals of the two measurements have the same weight.

To avoid local minima affecting the fit process, linked to the strong nonlinear character of the Nekhoroshev-like diffusion coefficient, the fit was performed with an initial scan over the $\kappa$ parameter, in line with the approach used in Ref.~\cite{bazzani2020diffusion}. Similarly to the previous work, a step of $0.01$ was considered for such a scan. For each $\kappa$ considered, the parameters $I_\ast$ and $I_0$ were initially scanned by brute force over a fixed range of values. The $I_\ast$ and $I_0$ values that provided the best fit quality were then used as starting points for a final minimization of the residuals with the Levemberg-Marquardt algorithm. For each value of $I_0$, the value of $\rho(I_0)$ is evaluated a posteriori, considering the scraping measurement data at the lowest amplitude, the values of $\kappa$ and $I_\ast$ under consideration, and Eq.~\eqref{eq:equilibrium}.

The resulting scan over $\eta$ provides an overview of how the fit results change with the weights given to the two measurements and allows us to understand the robustness of the fit results and the reliability of the measurements themselves. Instead, the scan over $\kappa$ is performed with the goal of reducing the issues induced by the unavoidable correlation between $\kappa$ and $I_\ast$, which was also addressed in Ref.~\cite{Bazzani:2019csk}.

\section{Measurement results}\label{sec:results}

For the Beam~1 data, the diffusion coefficient was reconstructed considering both collimator scans, with the first scan combined with the second scraping and the second scan with the fourth scraping. 

For the Beam~2 data 
with BBCW on, the diffusion coefficient was reconstructed considering the first collimator scan and the second collimator scraping. As for the Beam~2 data with BBCW off, due to a very limited amount of data, the diffusion coefficient was reconstructed considering the reduced data available from the second collimator scan, without considering scraping measurement. For the last case, the value of $\rho(I_0)$ was inferred from the last scraping measurement performed on the third scraping. This adaptation of the fit procedure was performed to best use the available data and to provide an initial estimate of the diffusion coefficient for the case with BBCW off.  

For each fit performed, we present an overview of $\chi^2_{\mathrm{total}}$, as defined in Eq.~\eqref{eq:residuals}, for the chosen range of $\eta$ values. 
We then observe how the optimal value of $\kappa$ changes as a function of the different weights given to the two measurements, as well as the resulting values of $I_\ast$ and $I_0$ and the overall quality of the fit based on the case $\eta=1$.

\subsection{Results for Beam~1}

\begin{figure}[thp]
\centering
    \includegraphics[width=0.8\columnwidth]{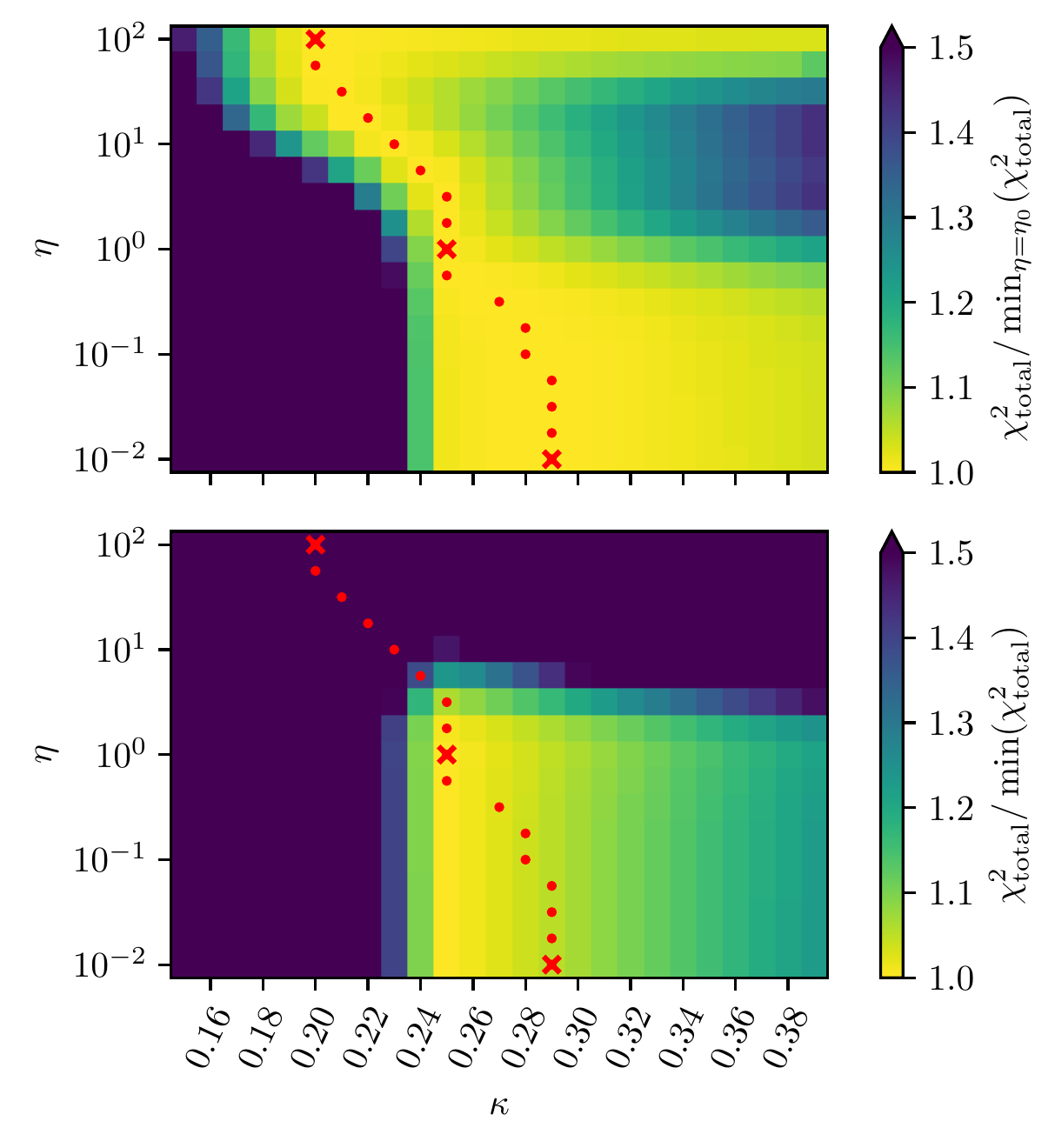}
    \caption{Behavior of $\chi^2_{\mathrm{total}}$ for the two measurements of Beam~1. The residuals are shown as a function of $\eta$ and $\kappa$, ranging from an extreme weight to the scan data to an extreme weight to the scraping data. In the top plot, $\chi^2_{\mathrm{total}}$ is evaluated for a certain $\eta$ value, and a normalization is performed to the minimum of $\chi^2_{\mathrm{total}}$ for that specific $\eta$ value. In the bottom plot, $\chi^2_{\mathrm{total}}$ is always evaluated for $\eta=1$, and the normalization is performed to the minimum value observed for the entire range of explored configurations. The red dots indicate the minimum value observed for each $\eta$ of the top plot, and are similarly reported in the bottom plot. In the bottom plot, the absolute minimum corresponds to the cross in the $\eta=1$ row.} 
    \label{fig:residuals-b1}
\end{figure}

Figure~\ref{fig:residuals-b1} reports the evolution of $\chi^2_{\mathrm{total}}$ for the two measurements of Beam~1. 
When evaluating $\chi^2_{\mathrm{total}}$ as a function of $\eta$, we obtain values that cannot be directly compared, due to the different numerical orders of magnitude obtained from the residuals of the two measurements. To recover a valid overview of the fit behavior, we normalize the values of $\chi^2_{\mathrm{total}}$ for each value of $\eta$ to the minimum value observed $\min(\chi^2_{\mathrm{total}})$ for that specific value of $\eta$. In doing so, we can clearly observe the evolution of $\chi^2_\mathrm{total}$ for the different values of $\eta$ and detect the value of $\kappa$ corresponding to the minimum. A similar optimal value of $\kappa$ is observed for $\eta=10^{-2}$ and $\eta=1$, namely around $0.29$, while a lower value of $0.19$ is preferred for $\eta=10^2$.

An overview of $\chi^2_{\mathrm{total}}$, evaluated for $\eta=1$, is also presented for the various choices of $\kappa$ and $\eta$ in Fig.~\ref{fig:residuals-b1}, where it is possible to observe how an optimal value for $\kappa=0.25$ is maintained.

\begin{figure*}[thp]
    \includegraphics[width=\textwidth]{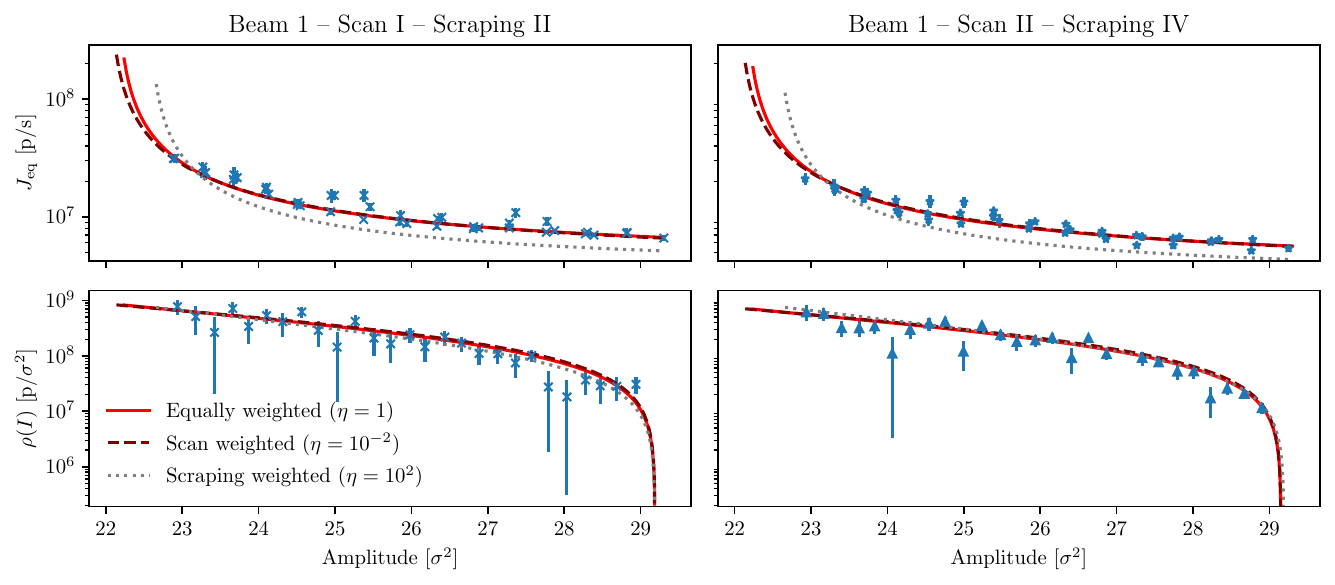}
    \caption{Best fits for Beam~1 for three choices of $\eta$, namely, preferring the scan data, preferring the scraping data, and the equally weighted setting. The fit that gives more weight to the scraping data utterly fails to replicate the observed equilibrium current. In contrast, the fit that gives more weight to the scan data accurately represents the reconstructed beam-halo population. Additionally, the equally weighted setting closely matches the fit that emphasizes the scan data.}
    \label{fig:fit-b1}
\end{figure*}

\begin{figure}[thp]
\centering
    \includegraphics[width=0.8\columnwidth]{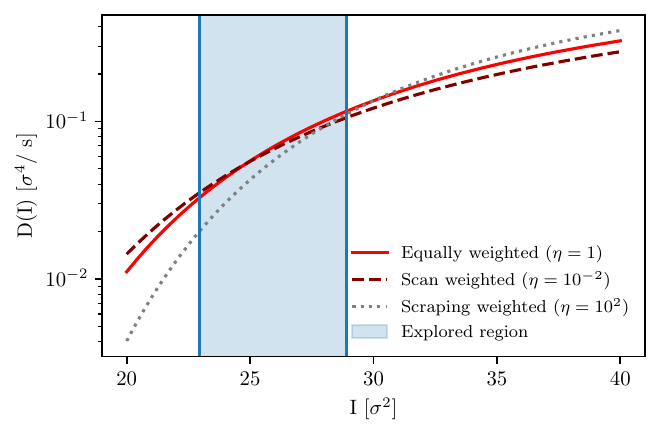}
    \caption{Reconstructed diffusion coefficient for Beam~1 for three choices of $\eta$, namely, preferring the scan data ($\eta=10^{-2}$), preferring the scraping data ($\eta=10^{2}$), and equally weighting them ($\eta=1$).}
    \label{fig:diffusion-b1}
\end{figure}

Three best fits for three choices of $\eta$, namely, $\eta=10^{-2}$, $\eta=1$, and $\eta=10^{2}$, are highlighted in Fig.~\ref{fig:residuals-b1} and presented in Fig.~\ref{fig:fit-b1}, with the parameters reported in Table~\ref{tab:fit-b1}. It can be observed that the fit that gives more weight to the scraping data completely fails to reproduce the observed equilibrium current, while, conversely, the fit that gives more weight to the scan data also provides a good representation of the reconstructed beam-halo population. Furthermore, the equally weighted setting shows close agreement with the fit, further validating the approach that consists of giving more weight to the scan data. A visualization of the reconstructed diffusion coefficient for the three fits is shown in Fig.~\ref{fig:diffusion-b1}. All these considerations are also reflected in the $\chi^2_\text{total}$ evaluated with global normalization, presented in the bottom plot of Fig.~\ref{fig:residuals-b1}.

\begin{table}
\centering
    \caption{\label{tab:fit-b1}%
    Best fit results for Beam~1 for three choices of $\eta$.
    }
    \begin{tabular}{lccc} \hline
        &
        Large weight&
        Equal weight&
        Large weight\\
        & to scan&
        &
        to scraping\\
        &
        ($\eta=10^{-2}$)&
        ($\eta=1$)&
        ($\eta=10^{2}$)\\
        \hline
        \rule{0pt}{2ex}
        $\kappa$ & $0.29$ & $0.25$ & $0.19$ \\
        $I_\ast$ & $30.56\pm0.07$ & $30.10\pm0.08$ & $31.2\pm0.6$ \\ 
        $I_0$ & $22.06\pm0.05$ & $22.11\pm0.05$ & $22.7\pm0.5$ \\
        $\rho(I_0)$ ratio & $0.85\pm0.01$ & $0.84\pm0.02$ & $0.95\pm0.5$ \\ \hline 
    \end{tabular}
\end{table}

To perform a consistency check on the results obtained with this specific fit reconstruction and, more importantly, to confirm the coherence of the results with the general working hypothesis posed at the beginning of the work, we analyze the evolution of a Crank-Nicolson integration~\cite{crank1947practical} of Eq.~\eqref{eq:fp}, to check if the same behaviors and shapes can be observed for a full diffusive system and for one with a fixed source at the fitted position $I_0$, both with the main parameters given by those obtained for Beam~1 with $\eta=1$.

We consider an initial exponential distribution $\rho_0(I)=\exp(-I)$ (corresponding to a Gaussian distribution in physical coordinates), with the absorbing boundary condition set at the same initial amplitude of the collimator jaw used for the Beam~1 measurements. This initial distribution serves as a generic representation of the expected beam population.

\begin{figure}[thp]
    \centering
    \includegraphics[width=0.8\columnwidth]{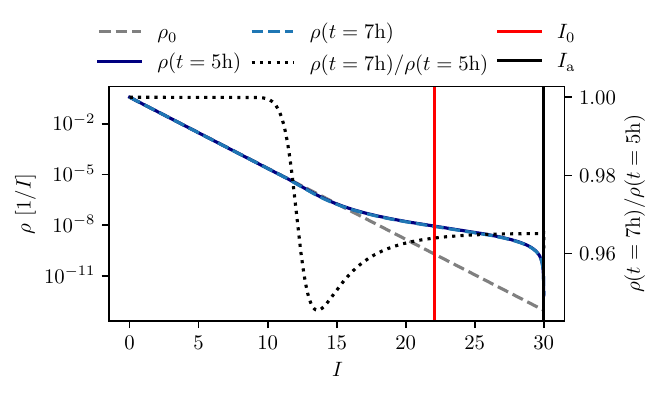}
    \caption{Evolution of the Crank-Nicolson integration of the Fokker-Planck equation using the parameters obtained from the fit for Beam~1 with $\eta=1$, listed in Table~\ref{tab:fit-b1}. The initial distribution is an exponential distribution with absorbing boundary condition set at the initial amplitude of the collimator jaw. The ratio between the distribution after eight hours of time and the distribution after ten hours of time is also shown.}
    \label{fig:crank-1}
\end{figure}

During the previous operations, which lasted more than 5 hours, Beam~1 went through four different changes in the crossing angle, and a total 20\% reduction in beam intensity. This implies that the system evolved under a time-dependent diffusion coefficient, which is outside of the theoretical framework that we have developed. Therefore, these effects cannot be included in numerical simulations. To evaluate the system's behavior under a generic Nekhoroshev-like diffusive process and test our working hypothesis, we use the diffusion coefficient obtained by the fitting procedure with $\eta=1$. We simulate the system's evolution over four hours of machine time. The resulting distribution, shown in Fig.~\ref{fig:crank-1}, matches the expected pattern from Ref.~\cite{our_paper9}, displaying a core region where the initial distribution is unaffected by the diffusion processes and a tail region where the distribution transitions to the expected $\rho_\mathrm{eq}(I)$ shape.

The system is then integrated for an additional two hours, with the goal of analyzing the evolution of the $\rho(I)$ distribution, especially at the amplitude $I_0$. The resulting distribution is also reported in Fig.~\ref{fig:crank-1}, along with the ratio between the two distributions. The evolution after the extra two hours of time shows a decrease in the beam-halo population at the amplitude $I_0$ of the order of $0.96$, slightly higher than the ratio obtained from the fit, but still in reasonable agreement with the fit results (i.e.~$0.84\pm 0.02$). The higher ratio observed in the simulation can be reconciled with the measured one by considering that in the numerical integration of the Fokker-Planck equation other sources of intensity loss are not included, while in the real machine, there is also burn-off induced by the two beams being colliding.

\begin{figure}[thp]
    \centering
    \includegraphics[width=0.8\columnwidth]{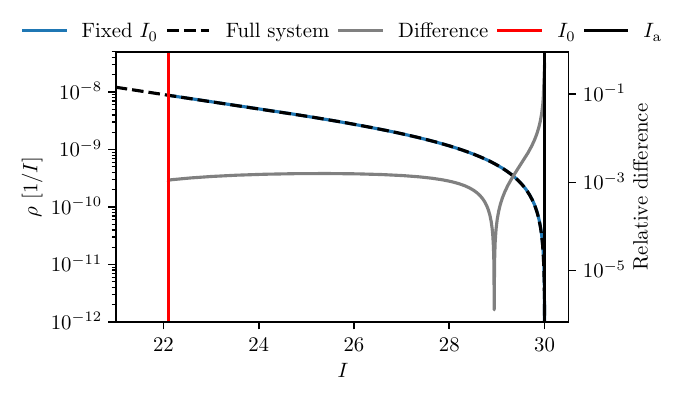}
    \caption{Comparison of the Crank-Nicolson integration of the Fokker-Planck equation using the parameters obtained from the fit for Beam~1 with $\eta=1$, available in Table~\ref{tab:fit-b1}, with the Crank-Nicolson integration of the Fokker-Planck equation using the same parameters, but with a fixed source term at the amplitude $I_0$. The relative error between the two distributions is also shown.}
    \label{fig:crank-2}
\end{figure}

To validate our hypothesis about using a fixed source term at amplitude $I_0$, we compare the distribution shapes obtained from two Crank-Nicolson simulations: one of the entire system after eight hours and one with a fixed source term at amplitude $I_0$, set to the value from the initial integration. As shown in Fig.~\ref{fig:crank-2}, the distributions agree closely, with a relative error around $10^{-3}$. This indicates that the fixed source term at amplitude $I_0$ is a valid approximation and that the diffusion process primarily drives the evolution of the beam-halo population.

\subsection{Results for Beam~2}

\begin{figure}[thp]
\centering
    \includegraphics[width=0.8\columnwidth]{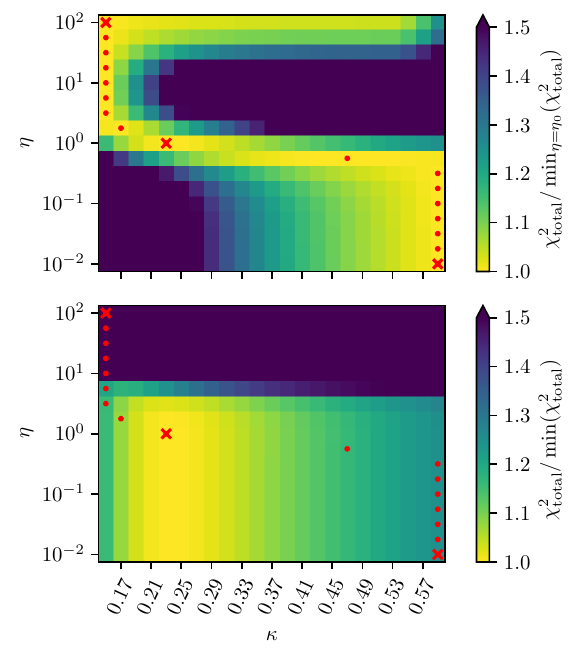}
    \caption{Overview of the $\chi^2_{\mathrm{total}}$ for the scan and scraping measurement of Beam~2 with BBCW on, following the same overview provided in Fig.~\ref{fig:residuals-b1}. Residuals are shown as a function of $\eta$ and $\kappa$, ranging from an extreme weight to the scan data to an extreme weight to the scraping data. In the top plot, $\chi^2_{\mathrm{total}}$ is evaluated for a certain $\eta$ value, and a normalization is performed to the minimum of $\chi^2_{\mathrm{total}}$ for that specific $\eta$ value. In the bottom plot, $\chi^2_{\mathrm{total}}$ is always evaluated for $\eta=1$, and the normalization is performed to the minimum value observed for the entire range of explored configurations. The red dots indicate the minimum value observed for each $\eta$ of the top plot, and are similarly reported in the bottom plot. In the bottom plot, the absolute minimum corresponds to the cross in the $\eta=1$ row.}
    \label{fig:residuals-b2}
\end{figure}

\begin{figure}[thp]
\centering
    \includegraphics[width=0.8\columnwidth]{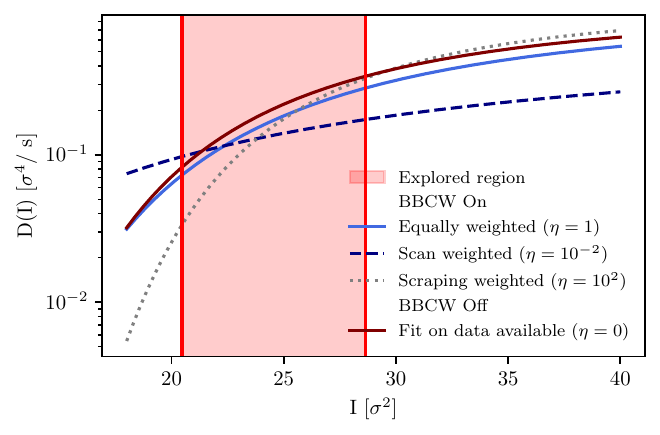}
    \caption{Reconstructed diffusion coefficient for Beam~2 for BBCW on and BBCW off. The three choices of $\eta$ are considered for the case of BBCW on, while a setting equivalent to $\eta=0$ is considered for the case of BBCW off. The diffusion coefficient for the case of BBCW off is observed to be greater than most of the reconstructions obtained for the case of BBCW on.}
    \label{fig:diffusion-b2}
\end{figure}

\begin{figure*}[thp]
    \includegraphics[width=\textwidth]{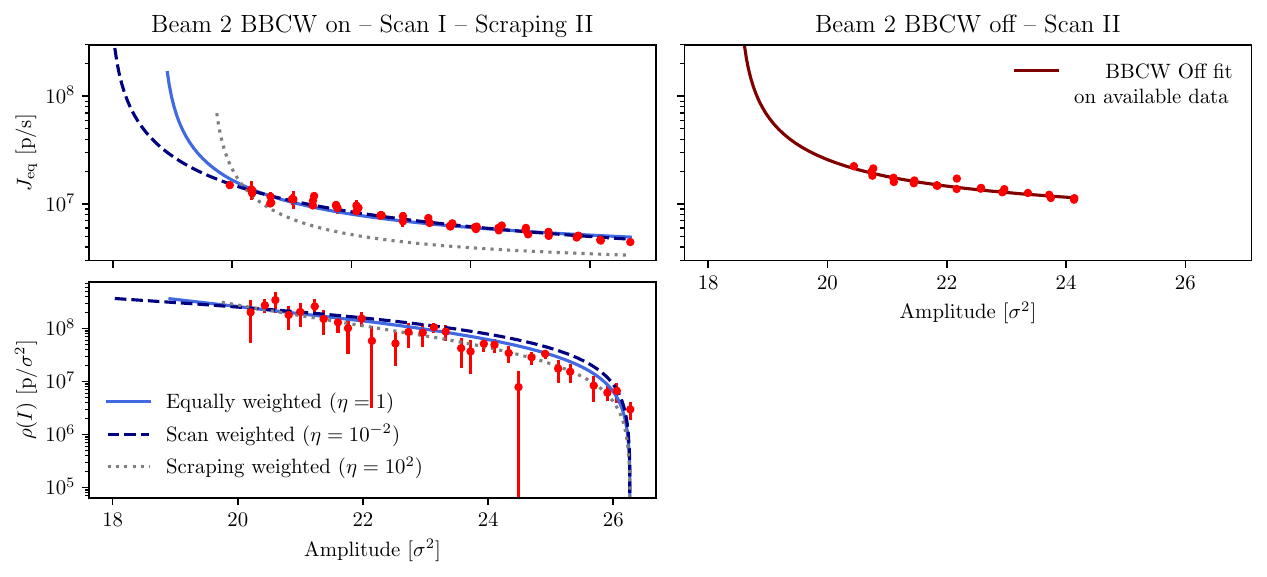}
    \caption{Best fits for Beam~2 for BBCW on (top two plots) and available data for BBCW off (bottom plot). The three choices of $\eta$ are considered for the BBCW on case, while a setting corresponding to $\eta=0$ is considered for the BBCW off case.} 
    \label{fig:fit-b2}
\end{figure*}

The $\chi^2_{\mathrm{total}}$ for the scan and scraping measurement of Beam~2 with BBCW on are shown in Fig.~\ref{fig:residuals-b2}, where the residuals are shown for a range of $\eta$ and $\kappa$ values. 
Unlike what was observed for the Beam~1 case, we see a saturation of $\kappa$ to the minimum value explored. 
This result is a strong symptom of low information content in the scan data and, as can be observed both in Table~\ref{tab:fit-b2} and in Fig.~\ref{fig:diffusion-b2}, the diffusion coefficient reconstructed from the various strategies appears to vary significantly as a function of the possible choice for the value of $\eta$. The best fits for Beam~2 with BBCW on are shown in Fig.~\ref{fig:fit-b2} (top plots).

The small value of the diffusion coefficient results in the consequent lack of a clear nonlinear character in the evolution of the equilibrium current during the collimator scan. This, in turn, prevents the fit from providing a reliable choice for the parameter $\kappa$. Inevitably, this implies that a deeper collimator scan would have been required. Despite this limitation, it can be observed that the equally weighted setting, with $\eta=1$, provides a good representation of both the equilibrium current and the beam-halo population, with a non-saturated value of $\kappa$. The results of the fit for $\eta=1$ highlight a value of $\kappa$ of $0.23$, which is close to the values of $\kappa$ observed for Beam~1.

As for the case of Beam~2 with BBCW off, 
the fit reconstruction performed corresponds in our notation to a value of $\eta=0$. The best fit for Beam~2 with BBCW off is shown in the bottom plot of Fig.~\ref{fig:fit-b2}, with the parameters reported in Table~\ref{tab:fit-b2}. As a stronger nonlinear diffusion character is observed, the fit provides a more reliable estimate of the diffusion coefficient, with $\kappa=0.20$. 

A visualization and comparison of the reconstructed diffusion coefficient for Beam~2 with BBCW on and BBCW off is shown in Fig.~\ref{fig:diffusion-b2}. It is observed that the diffusion coefficient for the case of BBCW off is larger than most of the reconstructions obtained for the case of BBCW on, hinting to a stronger diffusion of the beam halo in the absence of the BBCW. All these results and considerations combined suggest that the BBCW is indeed effective in reducing the overall diffusion of the beam-halo population. 

\begin{table}[thp]
\centering
    \caption{\label{tab:fit-b2}%
    Best fit results for Beam~2. Available data for BBCW enables to test different choices of $\eta$, while the limited amount of data for BBCW off only enables to fit the scan data with a setting equivalent to $\eta=0$. Note that some values of $\kappa$ are saturated to the boundaries of the explored range.}
    \begin{tabular}{lccc|c}
    \hline
        &\multicolumn{3}{c|}{BBCW on}&BBCW off\\
        &
        Large weight&
        Equal weight&
        Large weight&
        Data available\\
        &
        to scan&
        &
        to scraping& \\
        &
        ($\eta=10^{-2}$)&
        ($\eta=1$)&
        ($\eta=10^{2}$)&
        ($\eta=0$)\\
        \hline
        \rule{0pt}{2ex}
        $\kappa$ & $0.59$ (max.) & $0.23$ & $0.15$ (min.) & $0.20$\\
        $I_\ast$ & $24.51\pm0.13$ & $23.17\pm0.08$ & $23.9\pm0.4$ & $22.37\pm0.06$\\ 
        $I_0$ & $17.93\pm0.09$ & $18.81\pm0.09$ & $19.6\pm1.0$ & $18.51\pm0.07$\\ \hline
    \end{tabular}
\end{table}

\section{Conclusions and outlook}\label{sec:conclusions}

In this paper, we have presented and discussed in detail a general framework that has been recently applied to beam measurements of the Nekhoroshev-like diffusion coefficient characterizing the transverse beam-halo dynamics in the LHC. These measurements were performed using scans and scraping with the jaw of a collimator, using both LHC counter-rotating proton beams. These results show remarkable agreement with the behavior described by our diffusive framework. The novelty of the approach used in this work lies in the successful combination of beam-halo population and beam-loss measurements to reconstruct a general diffusion coefficient. The ultimate goal of this approach is to use the resulting diffusion coefficient to perform numerical simulations using a Fokker-Planck equation to study the evolution of the beam-halo population. This approach has outstanding advantages over standard numerical tracking tools, particularly in the level of detail with which the beam-halo evolution can be determined and, more importantly, in the time scales over which the evolution can be considered. 

During the experimental session, the long-range beam-beam wire compensator for Beam~2 was used. This provided a unique opportunity to probe the impact of this device on the beam-halo dynamics. In fact, our findings demonstrate that the diffusion coefficient of the beam-halo population is significantly influenced by the presence of the BBCW. For the first time, we have observed a reduction in the diffusion coefficient when the BBCW is switched on. We plan to study the potential benefit of this observed effect on the operation of HL-LHC. 

Looking ahead, future beam measurements can benefit from incorporating collimator scrapings at lower transverse amplitudes. This would provide additional data points and improve the accuracy of the estimation of the diffusion coefficient, especially in low-diffusion regimes such as those observed with BBCW on. Furthermore, faster and larger collimator steps, e.g., around \SI{10}{\micro m}, could be used with the goal of increasing signal strength and reducing the potential interference of out-of-equilibrium relaxation during scrapings. By implementing these improvements, future analyzes in the LHC and similar circular accelerators can enhance the understanding of beam dynamics and improve the reliability of diffusion coefficient measurements.

\section*{Acknowledgements}

The authors would like to thank S. M. Vigo and B.~Salvachua for their support in providing the calibration of the BLM data used in this work. The authors would also like to thank B. Lindstrom and M. D'Andrea for the support provided during the data taking.

\appendix

\section{Considerations on Beam 2 data and single-bunch instabilities}\label{sec:b2}

During the measurements on Beam~2 with BBCW off, notable single-bunch instabilities were detected, affecting the quality of the data. These instabilities manifested as abrupt spikes in the BLM loss signal, seemingly independent of concurrent collimator jaws adjustments. These spikes, significant enough to obscure the loss patterns under scrutiny, prompted a meticulous examination of the behavior of the individual bunches. Subsequently, data intervals heavily influenced by these instabilities were excluded from the analysis.

\begin{figure}[thp]
\centering
    \includegraphics[width=0.8\columnwidth]{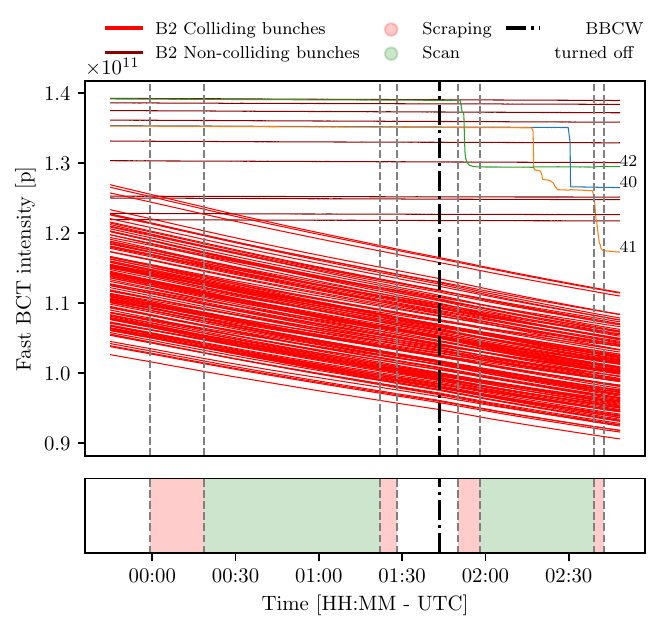}
    \caption{Time evolution of the bunch-by-bunch intensity measurement for Beam~2. Colliding and non-colliding bunches show an expected different trend in intensity evolution. The three non-colliding bunches affected by instabilities show sudden losses in intensity and are highlighted.}
    \label{fig:bunches_b2}
\end{figure}

As single-bunch instabilities were observed to affect only a limited number of non-colliding bunches, it was possible to inspect the fast BCT data to identify the bunches that were affected by the instabilities. As the fast BCT offers a high-resolution measurement of the bunch-by-bunch intensity, it was also possible to highlight when the most significant instabilities were observed. An overview of the bunch-by-bunch intensity for the duration of the measurement is shown in Fig.~\ref{fig:bunches_b2}, where it can be seen how three non-colliding bunches (namely, bunch slots 40, 41, 42) were affected by instabilities.

\begin{figure}[thp]
\centering
    \includegraphics[width=0.8\textwidth]{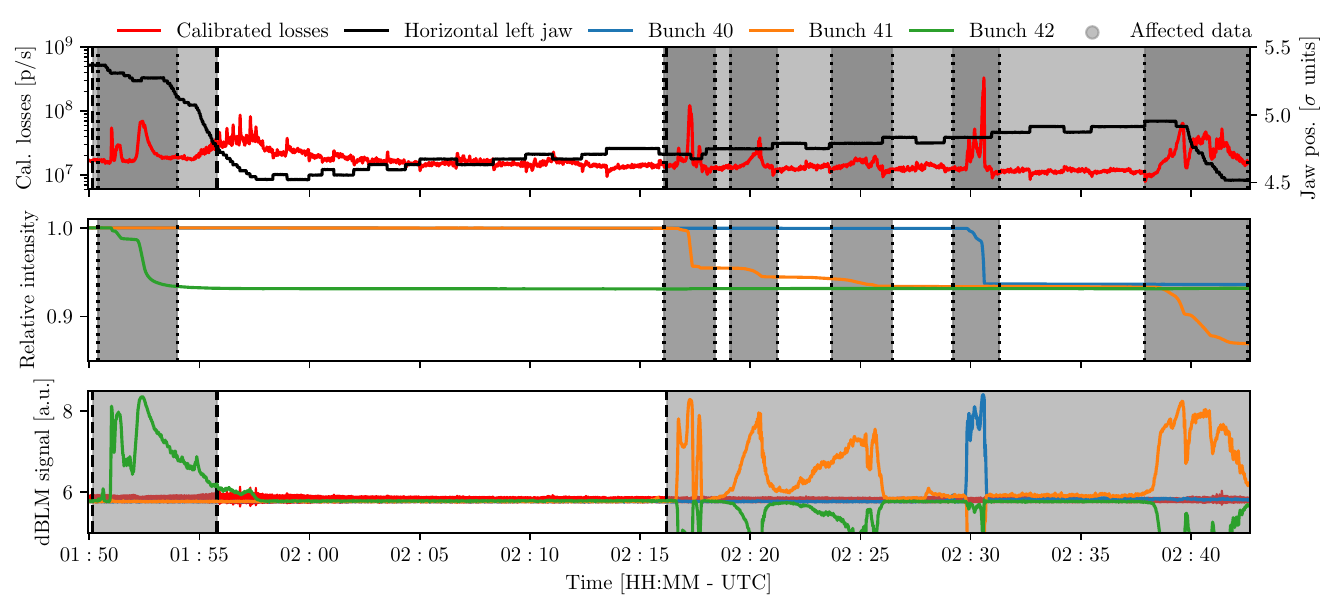}
    \caption{Overview of the excluded time windows for Beam~2 BLM loss data (top plot) with BBCW off. The time windows are highlighted in two shades of grey, with the darker shade indicating the regions where the three unstable bunches manifested strong losses in the fast BCT data (middle plot) and with the lighter shade indicating the regions where the dBLM signal (bottom plot) of the three unstable bunches was higher than the signal of the colliding bunches.}
    \label{fig:b2_ugly}
\end{figure}

To further corroborate the observation of the single-bunch instabilities and to assess whether the different shape and emittance of the unstable bunches after partial losses could have affected the entirety of the measurement, an analysis of the dBLM histograms was performed. By comparing the signal of the three unstable bunches with the magnitude of the signal of the various colliding bunches, it was possible to observe how a higher signal was observed from the unstable bunches, even in regions where the fast BCT does not highlight a significant intensity loss, thus requiring the exclusion of an even larger portion of the data. A complete overview of both the bunch-by-bunch intensity and the dBLM data histograms for the three unstable bunches is shown in Fig.~\ref{fig:b2_ugly}.

The dBLM data are provided in arbitrary units and currently do not offer a direct conversion to the actual beam losses in \si{p/s}. Moreover, crosstalks and other artifacts are still present in the data and are not yet fully addressed, for example, it can be observed in Fig.~\ref{fig:b2_ugly} how the dBLM signal of bunch 42 also mirrors the signal of bunch 41, suggesting a crosstalk between the two bunches. Likewise for bunch 41 and bunch 40. Despite these limitations, we can still rely on the dBLM data to provide a precise overview of the bunch-by-bunch losses to guide us to the exclusion of the data affected by the instabilities.

By finally excluding the time windows in which any of the three unstable bunches manifested a higher signal in the dBLM loss signal compared to the various colliding bunches, we were able to provide a reliable data set for the diffusion fit. The excluded time windows are highlighted in Fig.~\ref{fig:b2_ugly}, and also reported in Fig.~\ref{fig:collimator_scan}.

\clearpage
\bibliographystyle{unsrt}
\bibliography{bib}

\end{document}